\documentclass[journal,12pt,onecolumn,draftclsnofoot,]{IEEEtran}

\usepackage[noadjust]{cite}

\usepackage{amsmath}

\usepackage{algpseudocode}

\usepackage{epsfig}

\usepackage[linesnumbered,ruled,vlined]{algorithm2e}

\usepackage{graphicx}

\usepackage{tabularx}

\usepackage{adjustbox}

\usepackage{multirow}

\usepackage{array}

\usepackage{subcaption}

\usepackage{float}

\usepackage{fixltx2e}

\usepackage{dblfloatfix}

\usepackage{url}

\begin{document}

\title{An unsupervised deep learning framework for medical image denoising}

\author{Swati Rai, Jignesh S. Bhatt,
        and S. K. Patra,~\IEEEmembership{Senior Member,~IEEE.}\\
        Indian Institute of Information Technology Vadodara, India}
        


\maketitle
\begin{abstract}
Medical image acquisition is often intervented by unwanted noise that corrupts the information content. This paper introduces an unsupervised medical image denoising technique that learns noise characteristics from the available images and constructs denoised images. It comprises of two blocks of data processing, viz., patch-based dictionaries that indirectly learn the noise and residual learning (RL) that directly learns the noise. The model is generalized to account for both $2D$ and $3D$ images considering different medical imaging instruments. The images are considered one-by-one from the stack of MRI/CT images as well as the entire stack is considered, and decomposed into overlapping image/volume patches. These patches are given to the patch-based dictionary learning to learn noise characteristics via sparse representation while given to the RL part to directly learn the noise properties. K-singular value decomposition (K-SVD) algorithm for sparse representation is used for training patch-based dictionaries. On the other hand, residue in the patches is trained using the proposed deep residue network. Iterating on these two parts, an optimum noise characterization for each image/volume patch is captured and in turn it is subtracted from the available respective image/volume patch. The obtained denoised image/volume patches are finally assembled to a denoised image or $3D$ stack. We provide an analysis of the proposed approach with other approaches. Experiments on MRI/CT datasets are run on a GPU-based supercomputer and the comparative results show that the proposed algorithm preserves the critical information in the images as well as improves the visual quality of the images.
\end{abstract}


\section{Introduction}

\IEEEPARstart{N}{oise} is the unwanted energy which is mixed during the acquisition, transmission, and/or reconstruction of an image. Though the noise cannot be altogether eliminated, however, it can be reduced at acquisition time. Post-processing of acquired imagery using data processing algorithms is used to reduce its effects. In such applications, denoising is a major challenge for the researchers \cite{andrews1977digital, buades2005review, talebi2013global, toutain2015unified, ker2017deep, purwins2019deep, ravani2019, choi2020statnet}. Denoising is an inverse ill-posed problem \cite{hansen1998rank} which is classically addressed by specifying a forward model and then invert it for the unknowns \cite{tirer2020back}. Recent developments are exploring the use of deep learning techniques for the denoising \cite{ker2017deep, jin2017deep, wurfl2018deep, han2019image, shende2019brief}. 

\par Denoising is the fundamental step in medical image processing applications \cite{sharma2018early, deshpande2019bayesian} while doctors and medical practitioners most often rely on these processed images for the diagnosis. In particular, magnetic resonance imaging (MRI) and computed tomography (CT) scans are used to observe the internal structure as well as any defects like tumors or injuries present inside the body. Generally, MRI and CT images are affected by noise due to fluctuations in temperature of the scanner room, disturbance in the scanning machines and/or patient's movement during the image acquisition. Due to the noise, magnitude of the pixel/voxel values in the images/image stack are perturbed which leads to artifacts and loss of details in the images. It makes the diagnosis and disease prediction complicated.

\par The main considerations involved in medical image denoising algorithms include: a) edges in the denoised image should be preserved, i.e., filtering performed for denoising should not blur out the finer details of imagery and while at the same time, b) the visual quality of the denoised image should be preserved and improved. In this paper, we propose a novel unsupervised deep learning method using patch-based dictionary learning (DL) and residual learning (RL) in order to construct a dictionary-based deep residue network for denoising of MRI/CT images.   

\par Rest of the paper is arranged as follows: we begin with literature review in Section II. Section III explains the proposed approach for denoising $2D$ and $3D$ MRI/CT images. We also present a theoretical analysis of the proposed approach including algorithmic details. The results obtained after implementing our proposed model to the noisy MRI/CT datasets along with qualitative and quantitative comparisons with state-of-the-art are shown in Section IV. Finally, the paper is concluded in Section V with possible future direction.

\section{Related work}

Over the years, various medical image denoising methods have been proposed \cite{mohan2014survey, Kaur:2018:1573-4056:675, Diwakar2018, thanh2019, wen2020}. By and large, four broad philosophies are adopted: (a) filtering, (b) transformation, (c) statistical, and (d) learning-based methods. With the recent advances in computer technology and available resources, learning-based methods have gained a lot of attention. Hence, we review the learning-based approaches for denoising the MRI and CT images.   

\par The learning-based approaches can be further divided into three subcategories: supervised learning, semi-supervised learning, and unsupervised learning. In supervised learning, the model is trained with available data sets from which it can learn features called pre-learning or it can learn these features simultaneously during image reconstruction. It is found that the images are denoised using the supervised learning approach by incorporating wavelet transform (WT), curvelet transform (CuT), and optimization techniques in machine learning frameworks. The compressed sensing (CS) technique is used in denoising MRI images and called as CS-MRI. The CS is included with a dictionary learning approach to learn an overcomplete dictionary using k-singular value decomposition (K-SVD) method to give a sparse representation of an image \cite{aharon2006k}. CS-MRI is used to reconstruct MRI images consuming less acquisition time in a supervised way \cite{lustig2008compressed}. Again dictionary learning is used along with CS to reconstruct MRI images by training the model with denoised images \cite{ravishankar2010mr}. Subsequently, Bayesian approach is used with dictionary learning to denoise the MRI images \cite{huang2014bayesian}. Recently, the deep learning approach is explored with the classical methods to denoise the MRI images \cite{10.1007/978-3-030-00500-9_2}. Supervised learning is practiced to enhance the quality of CT and MRI images by removing noise and reducing the artifacts from them \cite{higaki2019improvement}. Very recently, directionality component is added to enhance the dictionary learning for MRI image reconstruction \cite{arun2020efficient}. 

\par A semi-supervised deep learning approach is used to reduce the noise from low-dose CT images without using original projection data by training the model with less number of denoised images \cite{chen2017low}. The low-dose CT images are mapped to their respective normal-dose part in a patch-by-patch manner using a deep convolutional neural network (CNN). Again, for low-dose CT images, a residual encoder-decoder CNN (RED-CNN) is formed by autoencoders and deconvolutional network which help in noise removal along with structural preservation and lesion detection \cite{chen2017ieee}. This uses normal-dose and low-dose CT images to train the network. Deep feed-forward CNN is then used to reduce noise from the images taking lesser number of clean images \cite{jifara2019medical}. This uses residual learning (RL) while batch normalization is used for regularization. Recently, generative adversarial network (GAN) is modified to Wasserstein GAN (WGAN) in order to denoise the MRI images in a semi-supervised manner \cite{ran2019}.  

\par It is a well-known fact that in medical imaging the availability of training dataset and ground truth is scare to train the model with supervised or semi-supervised settings. Therefore, a better approach is to investigate unsupervised learning models that can learn on their own only from the available images and could generate high-quality denoised images. A lot of attention is being given to low-dose CT as it reduces the risk on patients. To give promising results for CT images and to keep the crucial information intact, GAN is combined with perceptual similarity and Wasserstein distance using unsupervised learning \cite{yang2018low}. A deep neural network is recently trained in an unsupervised way using Poisson unbiased risk estimator (PURE) to denoise the low-dose CT image \cite{kim2020unsupervised}. 

\par Besides, denoising is an inverse ill-posed problem due to the existence of multiple solutions and inconsistency due to noise. Limited availability of labeled dataset in medical field makes it more complicated especially while solving using the learning-based approaches. Our approach in this paper is an unsupervised deep learning method that addresses the ill-posedness by learning the noise indirectly via learning the patch-based dictionaries (DL part) as well as residue (noise) is learned from available images (RL part). With the DL technique, we achieve sparse representation of the images. To this end, we choose orthogonal matching pursuit (OMP) to calculate the sparse coefficients and the K-SVD algorithm to update the patch-based $3D$ and $2D$ dictionaries from the images. On the other hand, RL part learns the residue, i.e. noise in our case, using the proposed deep residue network that comprises of convolution, rectified linear units (ReLU), and batch normalization layers along with carry forward connections facilitating the unsupervised training. With the knowledge of both sparse representation and residue, we obtain optimum residue for overlapping image/volume patches of data. These are finally used to denoise the input MRI/CT images. We discuss theoretical analysis, algorithmic aspects, and comparative experimental analysis with many state-of-the-art approaches using different MRI and CT image datasets. 


\section{Proposed Approach}

\begin{figure*}
\centering
\includegraphics[width=0.8\linewidth]{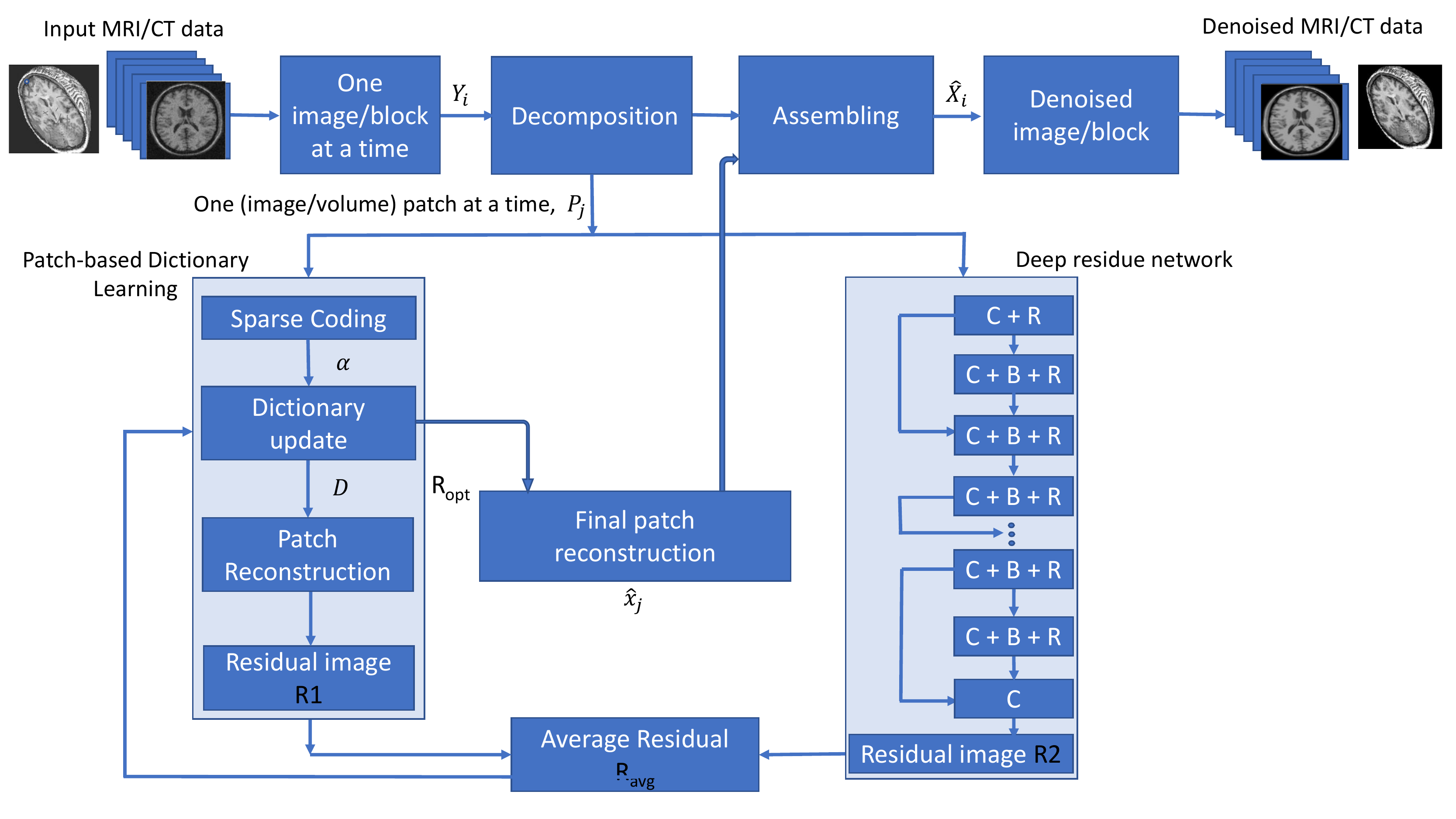}
\caption{Block diagram of the proposed unsupervised learning approach for MRI/CT denoising. C= Convolution, R= ReLU, and B= Batch normalization. All the functions are in $3D$ for voxel (block) processing and $2D$ for pixel (image) processing.}
\label{1}
\end{figure*}

In this section, we first define the problem and then discuss our proposed unsupervised learning approach for MRI and CT image denoising. Given medical MRI/CT $2D$/$3D$ image/stack, our objective is to estimate a corresponding denoised $2D$/$3D$ image/stack such that critical contents especially at edges in the estimated denoised images are preserved and visual information (quality) of the resultant (denoised) images is improved. In this work, we resort to the well-accepted data model for denoising medical images \cite{talebi2013global, gravel2004method, jifara2019medical},

\begin{equation}
    \mathbf{Y} = \mathbf{X} + \mathbf{Z},
\end{equation}
where $\mathbf{Y}$ is available (given) image, $\mathbf{X}$ is the corresponding denoised image (unknown), and $\mathbf{Z}$ is the noise. We conveniently consider that MRI images are corrupted by the Rician noise \cite{he2008nonlocal, aja2008noise} while CT images are corrupted by the Poisson noise \cite{ma2013dictionary, mascarenhas1993global}. Hence, given the $\mathbf{Y}$ image, our objective is to estimate denoised image $\mathbf{\widehat{X}}$ that is close to the $\mathbf{X}$, both qualitatively and analytically. Block diagram of the proposed approach is shown in Fig. 1. It mainly comprises of two parts: dictionary learning (DL) and residual learning (RL). The MRI and CT data are available in the form of a $3D$ image cube of internal body parts. We develop a model for both $3D$ and $2D$ processing of the MRI/CT data considering different generations of scanning machines. Note that while the proposed framework is generalized for $3D$ and $2D$ processing, however, user can perform either $3D$ block or $2D$ image for processing.

\par \textit{For $3D$ processing}: In our approach, we consider a block of images comprising of voxels. These images $\{\mathbf{Y}_i\}_{i=1}^{l}$ where $i$ is the index of an image and $l$ is the total images in the $3D$ cube. For voxel processing, each $3D$ block is of dimension $N \times N \times Q$ voxels. These images are first given to the decomposition stage. Here the $3D$ block is divided into overlapping block (volume) patches $\{\mathbf{P}_j\}_{j=1}^{r}$ each of size $n \times n \times q$ voxels, where $n << N$, and $j$ is the index of the block patch chosen from a total of $r$ block patches. These patches $\{\mathbf{P}_j\}_{j=1}^{r}$ are fed to the DL and RL parts for processing, again one block patch $\mathbf{P}_{j}$ at a time. 

\par \textit{For $2D$ processing}: Considering $3D$ volume data, we take one image at a time, each image has a dimension of $N \times N$ pixels and is decomposed into image patches of size $n \times n$ pixels. Now, these obtained patches $\{\mathbf{P}_j\}_{j=1}^{r}$ are given to the DL and RL parts for further processing, again one image patch $\mathbf{P}_{j}$ at a time. 

\par Referring to Fig. 1, there are three steps in the DL part: (a) sparse coding, (b) dictionary update, and (c) patch reconstruction. The role of DL part is to provide efficient representation of input MRI/CT so that, in turn, we have estimate of noise content via the sparse representation of information. This is an indirect way of learning noise characteristics. To start the DL process, we use an initial dictionary $\mathbf{D}_{init}$ of size $m \times k \times q$ for block processing and of size $m \times k$ for image processing, obtained using the discrete cosine transform (DCT). We consider an overcomplete dictionary since it has basis vectors greater than the dimension of the input patch vector, which allows to better capture underlying characteristics of the data. One may notice that for medical images, capturing the underlying information is vital for better processing and the final diagnosis. With the initial dictionary $\mathbf{D}_{init}$ and available patch $\mathbf{P}_j$, we first obtain the sparse coefficient $\boldsymbol{\alpha}_j$ of dimension $k \times 1 \times q$ for block processing and of $k \times 1$ for image processing, i.e., the sparse representation of a patch $\mathbf{P}_j$ is considered as:

\begin{equation}
    \mathbf{P}_j \approx \mathbf{D}_{init} \;   \boldsymbol{\alpha}_j,
\end{equation}

where the sparse coefficients of a image/block patch is computed using orthogonal matching pursuit \cite{tropp2007signal} as,

\begin{equation}
    \widehat{\boldsymbol{\alpha}}_j = \min_{\boldsymbol{\alpha}_j} \left(\frac{1}{2} ||\mathbf{P}_j - \mathbf{D}_{init} \boldsymbol{\alpha}_j||_{2}^{2} + \mu ||\boldsymbol{\alpha}_j||_0 \right).
\end{equation}
Here $\mu$ is the regularization parameter. We now estimate sparse dictionary $\mathbf{D}$ using the estimated sparse coefficients $\widehat{\boldsymbol{\alpha}}_j$,

\begin{equation}
  \mathbf{D} = \arg\min_{\mathbf{D}}{\sum_{j=1}^n ||\mathbf{P}_j - \mathbf{D} \widehat{\boldsymbol{\alpha}}_j||_{2}^{2}} \; \; \text{such that} \; \; ||\boldsymbol{\alpha}_j||_0 \leq s,
\end{equation}
where s is the sparsity. To this end, in order to update the dictionary, we employ K-SVD algorithm \cite{Aharonksvd2006}. Now this updated dictionary $\mathbf{D}$ and the estimated sparse coefficients $\widehat{\boldsymbol{\alpha}}_j$ are used to reconstruct denoised image/block patch $\widehat{\mathbf{X}}_j$ as:

\begin{equation}
     \widehat{\mathbf{X}}_j = \mathbf{D} \;   \widehat{\boldsymbol{\alpha}}_j.
\end{equation}

The residual patch $\mathbf{R1}_j$ can now be extracted by taking absolute difference of estimated denoised image/block patch $\widehat{\mathbf{X}}_j$ and available input image/block patch $\mathbf{P}_j$ as,

\begin{equation}
    \mathbf{R1}_j := |\mathbf{P}_j - \widehat{\mathbf{X}}_j|, \; \; \; \; \forall{j}.
\end{equation}

See that in equation (6) we have used absolute subtraction between the given patch and estimated denoised patch referring to our data model equation (1). Note that the residue patch $\mathbf{R1}_j$ consists of part of the noise contents due to representational limitations at the time of image acquisition. Thus, the proposed DL part indirectly learned the noise characteristics from the MRI/CT data. 

\par Now the residual learning (RL) part in the proposed model (Fig. 1) is designed to directly learn the noise characteristics present in the patches. As shown in Fig. 1, we pass the image/block patch $\mathbf{P}_j$ through the proposed deep residue network having depth $t$ comprises of the following layers: (a) First layer (C + R): C stands for the convolution process that is performed between a patch and a ﬁlter. Note that it will be $2D$ convolution for image processing while $3D$ convolution for block (volume) processing. Convolution helps to extract the features of the image/block and generate feature maps. In particular, $84$ ﬁlters of size $3 \times 3$ are employed that give rise to $84$ feature maps. Then rectiﬁed linear units (R) are used to introduce the non-linearity by using the $max(0,·)$ function. (b) Second layer to $(t-1)$ layer (C + B + R): Here, batch normalization (B) is introduced in between C and R. The B acts as a regularizer term and helps the network to use higher learning rates which in turn uplifts the denoising performance. Note that there are skip connections added in between alternate layers in deep residue network (Fig. 1). The layers having same dimension receive identity connection from the previous layer. While convolution layer is added in between the identity connection if dimension of recent input and previous input data is different. (c) Last layer (C): Finally convolution is performed to give the residual $\mathbf{R2}_j$ (noise) part learned from the input image/block patch. In this way we directly learn the noise from image/block MRI/CT data.

\begin{figure}[ht]
 \centering
 \includegraphics[width=0.7\linewidth,height=0.9\linewidth]{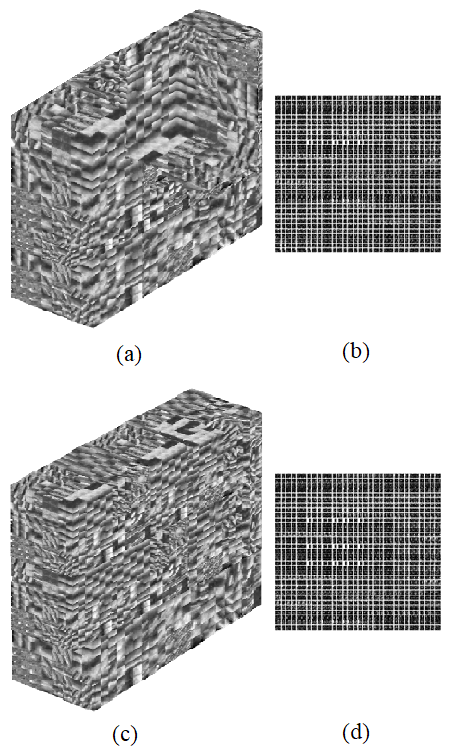}
 \caption{Obtained sample dictionary patches from real MRI \cite{ACRIN2020} and CT \cite{ACRIN2020} data: (a) from $3D$ MRI, (b) from $2D$ MRI, (c) from $3D$ CT, and (d) from $2D$ CT.}%
 \label{2}%
\end{figure}

\par Hence we now have $\mathbf{R1}_j$ residue using the DL, and $\mathbf{R2}_j$ residue using the RL. We construct residue $\mathbf{R}_{avg_j}$ by averaging $\mathbf{R1}_j$ and $\mathbf{R2}_j$, pixel-by-pixel, to preserve the noise characteristics learned by representation content (DL) and residue (RL). The averaged residue $\mathbf{R}_{avg_j}$ is then fed back to the DL stage in order to again update the estimated dictionary $\mathbf{D}$, as shown in Fig. 1:

\begin{multline}
 \mathbf{D} = \arg\min_{\mathbf{D}} \lambda {\sum_{j=1}^r ||\mathbf{P}_j - \mathbf{D} \boldsymbol{\alpha}_j||_{2}^{2}} + \mu ||\boldsymbol{\alpha}_j||_0 + \\ \frac{1}{r} ||\mathbf{R}_{avg_j} - \mathbf{R1}_j||_{F}^{2},
 \end{multline}

where $F$ is Frobenius norm and $\lambda$ is regularization parameter. Fig.2 displays sample dictionaries obtained for denoising both $3D$ and $2D$ MRI and CT data. Fig. 2(a) shows $3D$ dictionary of size $32 \times 32 \times 8$ and the size of $2D$ dictionary is $32 \times 32$ in Fig. 2(b) for MRI data. Similarly dictionaries for $3D$ and $2D$ CT data are shown in Fig. 2(c) and Fig. 2(d), respectively. Now the updated dictionary $\mathbf{D}$ is used to generate the optimum residue $\mathbf{R}_{opt}$ and give the final estimated denoised patch $\widehat{\mathbf{X}}_{opt_j}$ as: 

\begin{equation}
    \widehat{\mathbf{X}}_{opt_j} := |\mathbf{P}_j - \mathbf{R}_{opt_j}|, \; \; \; \;  \forall{j}.
\end{equation}

The entire process from equations (2) to (8) is repeated for all the $r$ image/block patches of an input image/block $\mathbf{Y}_i$. Finally estimated denoised patches are assembled to form an entire estimated denoised image/block $\widehat{\mathbf{X}}_i$ (Fig. 1). Note that the patches are overlapping, therefore, the voxels and pixels in the overlapping regions in the $\widehat{\mathbf{X}}_i$ are considered by local patch-level averaging. Finally, as shown in Fig. 1, the process is repeated for each image/block in the stack of MRI/CT images and estimate corresponding denoised stack of MRI/CT images. 

\subsection{Analysis of the proposed approach}

\par In this subsection, we analyse the proposed approach for its effectiveness and conduct analytical comparison with respect to state-of-the-art approaches. It is observed that noise is most often mixed in MRI and CT images during the image acquisition process. Therefore let us first basically understand how these medical images are acquired. In MRI, a patient's body is exposed to a very strong magnetic field, radio waves, and magnetic field gradients \cite{560324}. During the acquisition, both the frequency and phase of the MRI signals, called raw MRI data, are accumulated in a temporary image space and then inverse Fourier transform is computed to form a grayscale MRI image. It is found that in MRI, the probability density function (PDF) of noise follows the Rician distribution \cite{mohan2014survey}. Hence, referring to the data model in equation (1), one may write the conditional PDF of MRI data as,

\begin{equation}
        p_\mathbf{Y}(\mathbf{Y} | \mathbf{X}) = \frac{\mathbf{Y}}{\sigma^2} \;  e^{\frac{-(\mathbf{X}^2 + \mathbf{Y}^2)}{2 \sigma^2}} \; I_0 \left(\frac{\mathbf{X} \; \mathbf{Y}}{\sigma^2}\right),
\end{equation}
where $\mathbf{Y}$ is the image acquired having noise, $\sigma$ is the noise variance, $\mathbf{X}$ is the noiseless image intensity level (unknown), and $I_0(\cdot)$ is the zeroth order modified Bessel function used to induce smoothness in the curve. In CT scans, a thin beam of X-rays is passed through a patient's body from the source that is captured by the X-ray detectors, located opposite to the X-ray source \cite{GoodCT1979}. These signals are processed by the computer and cross-sectional images of the patient's body are generated. In CT images, the most common noise is the Poisson noise \cite{thanh2019}. This is mainly due to the usage of X-rays and scanning methods in the generation of the CT scans. The probability mass function (PMF) of CT data can thus be written in reference to the data model equation (1) as,

\begin{equation}
        p(\mathbf{Y} | \mathbf{X}) = \frac{e^{-\mathbf{X} t} (\mathbf{X} t)^\mathbf{Y}}{\mathbf{Y}!},
\end{equation}
where $\mathbf{Y}$ is the amount of photons (image intensities) measured over time interval $t$ by the sensor element, and $\mathbf{X}$ is the expected amount of photon (corresponding denoised image content) per unit time. It can be seen from equation (9) and (10) that both Rician and Poisson noise affect the magnitude of the MRI and CT images, respectively. 

\par Basically, MRI/CT images are perturbed mainly due to three major causes \cite{hoiting2005measuring, boas2012ct}: (a) Ambient temperature is not maintained inside the scanner room. Typical range of temperature required to maintain is $23 ^{\circ}$ to $24 ^{\circ}$ Celsius. Any variation beyond the said temperature range can cause artifacts in image. (b) Number of detectors used to capture the images. More number of detectors can reduce the scanning time, however, probability of noise is also increased. (c) Any movement of patient during scanning leads to artifacts and loss of finer details in the images. Besides, type of scanning machines impacts the image acquisition process. Third generation or below machines typically generate the $2D$ scans that can later be converted into $3D$ data while fourth and fifth generation machines directly provide $3D$ data as output. Hence, we have proposed an approach to perform denoising process on both $2D$ and $3D$ data.  

\par We observe that many researchers have explored the volumetric data procedure for the denoising medical images. In block-matching and 3D filtering (BM3D) \cite{dabov2007image}, the similar image patches are stacked to form $3D$ blocks and filtering is done on all the blocks. The inverse transform is then performed to get them back into $2D$ form. On the other hand, non-local means (NLM) \cite{manjon2008mri} processes the voxels by $2D$ filtering with a search and a neighborhood window. This is used to find out the similarity between the pixels and a parameter to control the degree of smoothness in an image. Underlying assumption is noisy patches will find the similarity with other patches containing noise. Therefore, as a side effect, the information present in the edges is lost and the edges become blur. In anisotropc diffusion filter (ADF) \cite{krissian2009noise}, voxels of the images are considered by combining domain and range Gaussian filters in order to find the geometric and photometric distances. The final estimated intensity value of a pixel is calculated by taking the average of the geometric and photometric distances among the pixels inside the selected spatial window. Hence, it enlarges the edge widths and makes them more blurry. Notably several recent approaches use $2D$ images for medical image denoising. Earlier K-SVD \cite{Aharonksvd2006} denoising is performed on individual images using the dictionary learning approach. Recently, RED-CNN \cite{chen2017ieee} takes the $2D$ images as input and combines autoencoders and deconvolutonal networks to preserve the image structures. More recently CNN-RL \cite{jifara2019medical} accepts the images in $2D$ form and instead of learning the mapping function of an image, it predicts the latent clean image. In total variation (TV) \cite{said2019total}, regularization is controlled in a way that more denoising process is applied in smooth regions and lesser at edge (discontinuity) regions of each image. In our proposed approach, we consider one block of $3D$ data at a time and construct dictionaries of overlapping block patches as well as learn noise from residual learning using the $3D$ blocks for a better denoising process. Considering different generations of scanning machines, $2D$ image processing is also included. We have generalized the proposed approach to consider each slice (image) of the MRI and CT data independently for denoising process (Fig. 1). 

\par It is found that patch-based methods effectively smoothens the homogeneous regions as well as preserves the finer details in an image. Our proposed model also learns patch-based dictionaries for each image/block from the set of input images. In TV \cite{said2019total}, the patches of an image are used in the edge detection scheme. When the TV norm of an image is too low it leads to over-smoothing and only edges are preserved if the norm is high. Therefore the approach \cite{said2019total} is sensitive to TV norm for denoising patches. The patch-based approach is also adopted in BM3D \cite{dabov2007image}. In K-SVD \cite{Aharonksvd2006} a similar approach to the K-means method is adopted, however, a single dictionary is learned for entire image. In the proposed approach, we are learning dictionaries for overlapping image/block patches of a dataset. Hence the critical information remains intact and edges are also preserved. 

\par In our approach, we learnt the noise characteristics from the image using proposed deep residue network. In CNN-RL \cite{jifara2019medical}, the residue is learned from the deep CNN layers and it is multiplied by a constant factor to normalize the elements. This normalized residue is then subtracted from the given image to form a denoised image. It may be noted this way the denoised image may observe the loss of information at the edges due to a single scaling factor to normalize the residue. Unlike it, our proposed unsupervised method considers the average of residue learned from dictionary learning and deep residue network. It essentially avoids the heuristic of hard coding of scaling factor. In RED-CNN \cite{chen2017ieee} the autoencoders and deconvolutional networks are used to preserve the structures in order to reduce noise from the images. This approach may over smoothed the edges as the data is compressed by the encoders and decoders. Our proposed approach uses the constraints to update the patch-based dictionaries and learn the residues that can better handle the ill-posed nature of the problem. The DL and RL parts work in a complementary manner so possible loss of information in one part is augmented by other part. Finally, the proposed model being unsupervised is practically useful since it only needs the available datasets and does not need clean (denoised) images as in the case of supervised training \cite{yang2018low, jifara2019medical, manjon2008mri, chen2017ieee}. 

\section{Results}

In this section, we evaluate our proposed unsupervised learning approach by conducting experiments on different MRI and CT images. We begin by providing details of the datasets, machine specifications, and parameter settings. We then compare and analyze the results obtained by our approach with state-of-the-art approaches.

\par \textit{Datasets}: We have used the datasets for MRI and CT images from the cancer imaging archive (TCIA) \cite{ACRIN2020}. It is an open-access database of medical images available for the research. In our experiment, data is consists of digital imaging and communications in medicine (DICOM) format. 

\par \textit{Machine specification}: All the algorithms are implemented in PARAM Shavak GPU-based supercomputer powered with two multicore CPUs, each with fourteen cores. It has NVIDIA GP100 accelerator card and 96 GB RAM. We have also used Intel Core i7-9750H CPU @ 2.60GHz with 20 GB RAM to generate the synthetic datasets for MRI and CT images. The programming is done using Python 3.7 and major libraries include matplotlib, skimage, numpy, scipy, pytorch, and pydicom. 

\begin{figure}[ht]
 \centering
 \includegraphics[width=0.7\linewidth,height=0.8\linewidth]{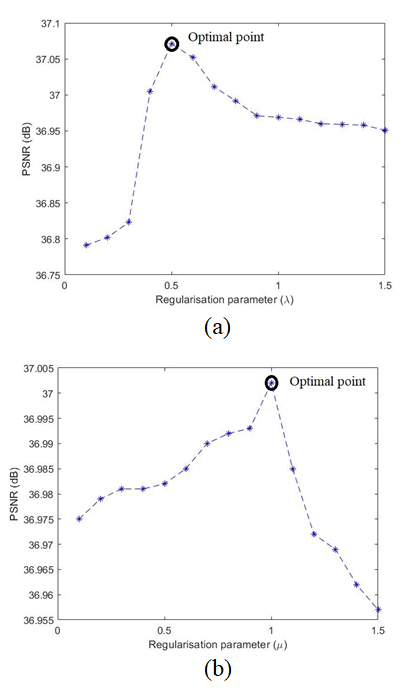}
 \caption{Sensitivity analysis of regularization parameters: (a) for $\lambda$, and (b) for $\mu$ in equation (7).}%
 \label{3}%
\end{figure}

\begin{figure*}
\centering
\includegraphics[width=0.8\linewidth]{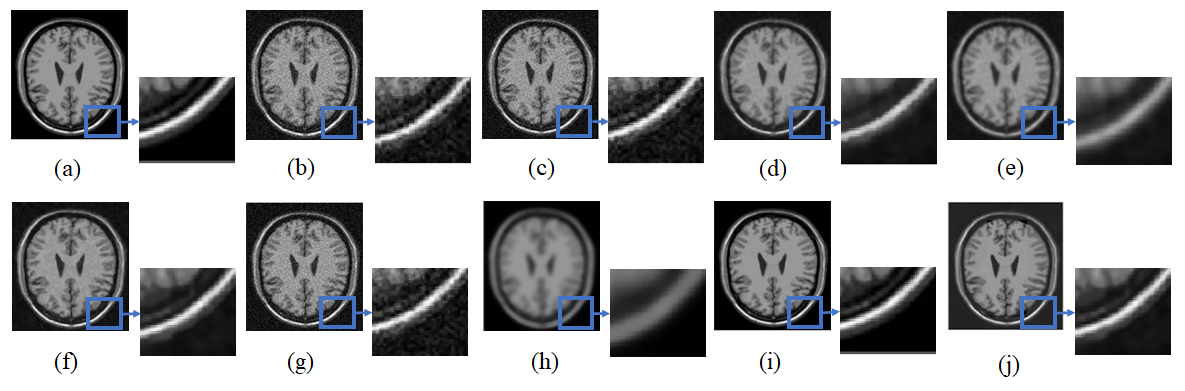}
\caption{Denoising by different algorithms by adding $5 \%$ Rician noise in MRI image \cite{ACRIN2020}: (a) Ground truth, (b) CNN-RL \cite{jifara2019medical}, (c) RED-CNN \cite{chen2017ieee}, (d) K-SVD \cite{Aharonksvd2006}, (e) TV \cite{said2019total}, (f) BM3D \cite{dabov2007image}, (g) NLM \cite{manjon2008mri}, (h) ADF \cite{krissian2009noise}, (i) Proposed $3D$, and (j) Proposed $2D$.}
\label{4}
\end{figure*}

\begin{figure*}
\centering
\includegraphics[width=0.8\linewidth]{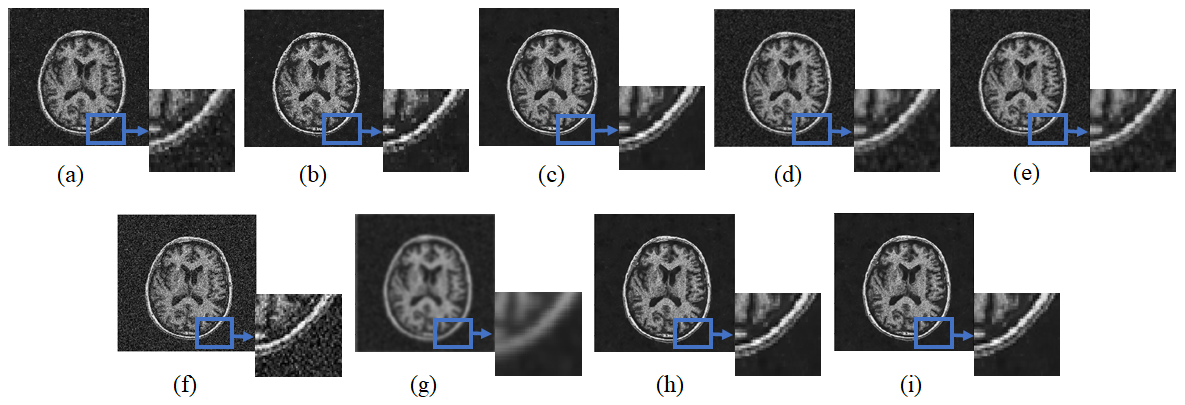}
\caption{Denoising by different algorithms on real MRI brain scans \cite{ACRIN2020}: (a) CNN-RL \cite{jifara2019medical}, (b) RED-CNN \cite{chen2017ieee}, (c) K-SVD \cite{Aharonksvd2006}, (d) TV \cite{said2019total}, (e) BM3D \cite{dabov2007image}, (f) NLM \cite{manjon2008mri}, (g) ADF \cite{krissian2009noise}, (h) Proposed $3D$, and (i) Proposed $2D$.}
\label{5}
\end{figure*}

\begin{figure*}
\centering
\includegraphics[width=0.8\linewidth]{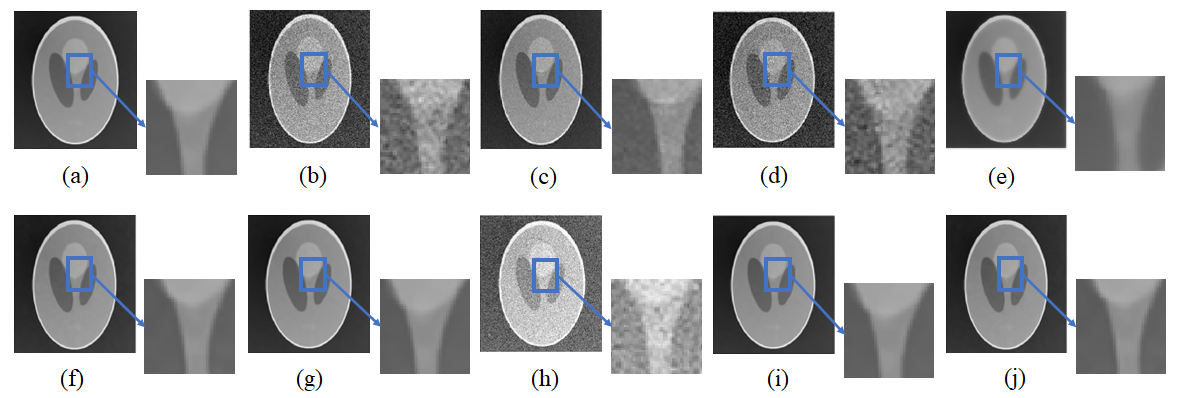}
\caption{Denoising by different algorithms by adding $5 \%$ Poisson noise to the Shepp-logan CT image \cite{ACRIN2020}: (a) Ground truth, (b) CNN-RL \cite{jifara2019medical}, (c) RED-CNN \cite{chen2017ieee}, (d) K-SVD \cite{Aharonksvd2006}, (e) TV \cite{said2019total}, (f) BM3D \cite{dabov2007image}, (g) NLM \cite{manjon2008mri}, (h) ADF \cite{krissian2009noise}, (i) Proposed $3D$, and (j) Proposed $2D$.}
\label{6}
\end{figure*}

\begin{figure*}
\centering
\includegraphics[width=0.8\linewidth]{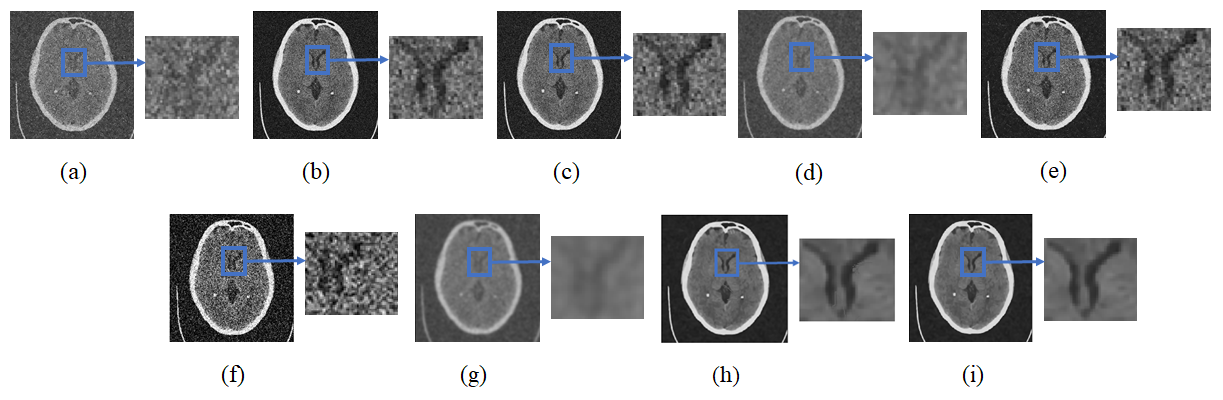}
\caption{Denoising by different algorithms on real CT brain scans \cite{ACRIN2020}: (a) CNN-RL \cite{jifara2019medical}, (b) RED-CNN \cite{chen2017ieee}, (c) K-SVD \cite{Aharonksvd2006}, (d) TV \cite{said2019total}, (e) BM3D \cite{dabov2007image}, (f) NLM \cite{manjon2008mri}, (g) ADF \cite{krissian2009noise}, (h) Proposed $3D$, and (j) Proposed $2D$.}
\label{7}
\end{figure*}


\begin{table}
\resizebox{\textwidth}{!}{
\begin{tabular}{|c|c|c|c|c|c|c|c|c|c|c|c|}
\hline
\multirow{2}{*}{\textbf{Measures}}                                            & \multirow{2}{*}{\textbf{\begin{tabular}[c]{@{}c@{}}Ideal \\  value\end{tabular}}} & \multirow{2}{*}{\textbf{\begin{tabular}[c]{@{}c@{}}Rician noise\\ levels\end{tabular}}} & \multicolumn{9}{c|}{\textbf{Methods}}                                                                                                                                                                                                                                                      \\ \cline{4-12} 
                                                                              &                                                                                   &                                                                                          & {\begin{tabular}[c]{@{}c@{}}\textbf{CNN-RL}\\\cite{jifara2019medical}\end{tabular}} & 
                                    \multicolumn{1}{l|}{{\begin{tabular}[c]{@{}c@{}}\textbf{RED-CNN}\\\cite{chen2017ieee}\end{tabular}}} &
                                    \multicolumn{1}{l|}{{\begin{tabular}[c]{@{}c@{}}\textbf{K-SVD}\\\cite{Aharonksvd2006}\end{tabular}}} &
                                    \multicolumn{1}{l|}{{\begin{tabular}[c]{@{}c@{}}\textbf{TV}\\\cite{said2019total}\end{tabular}}} &
                                    
                                    \multicolumn{1}{l|}{{\begin{tabular}[c]{@{}c@{}}\textbf{BM3D}\\\cite{dabov2007image}\end{tabular}}} &
                                    {\begin{tabular}[c]{@{}c@{}}\textbf{NLM}\\\cite{manjon2008mri}\end{tabular}} &
                                    {\begin{tabular}[c]{@{}c@{}}\textbf{ADF}\\\cite{krissian2009noise}\end{tabular}} & 
                                    \multicolumn{1}{l|}{{\begin{tabular}[c]{@{}c@{}}\textbf{Proposed}\\\textbf{3D}\end{tabular}}} &
                                    \multicolumn{1}{l|}{{\begin{tabular}[c]{@{}c@{}}\textbf{Proposed}\\\textbf{2D}\end{tabular}}} \\ \hline
\multirow{3}{*}{\begin{tabular}[c]{@{}c@{}}\textbf{PSNR} (dB)\\\cite{hore2010image}\end{tabular}} & \multirow{3}{*}{High}                                                             & 5 \%                                                                                    & 33.685          & 35.663 & 33.252        & 28.928    & 36.213    & 29.166       & 28.461                                                   & \textbf{42.992}        & \textbf{39.421}                           \\ \cline{3-12} 
                                                                              &                                                                                   & 10 \%                                                                                         & 33.462          & 34.673     & 32.336   & 28.033     & 35.121   & 28.872       & 28.124                                                    & \textbf{41.870}        & \textbf{37.071}                           \\ \cline{3-12} 
                                                                              &                                                                                   & 15 \%                                                                                   & 32.747          & 31.982     & 31.032   & 27.781   & 28.164   & 34.929      & 28.022                                                                                & \textbf{40.297}            & \textbf{35.185}                                            \\ \hline
\multirow{3}{*}{\textbf{SSIM} \cite{hore2010image}}      & \multirow{3}{*}{1}                                                                & 5 \%                                                                                    & 0.801           & 0.824     & 0.809     & 0.677  & 0.832         & 0.721        & 0.708                                                         & \textbf{0.899}         &  \textbf{0.881}                       \\ \cline{3-12} 
                                    &                                                                                   & 10 \%                                                                                   & 0.791           & 0.813   & 0.803     & 0.656    & 0.822      & 0.709        & 0.672                                                                & \textbf{0.893}          & \textbf{0.869}                        \\ \cline{3-12} 
                                    &                                                                                   & 15 \%                                                                                   & 0.724           & 0.762              & 0.736        & 0.573     & 0.792     & 0.683        & 0.618                                                      & \textbf{0.874}    & \textbf{0.826}                              \\ \hline
\multirow{3}{*}{\textbf{RMSE} \cite{sara2019image}}      & \multirow{3}{*}{0}                                                                & 5 \%                                                                                    & 19.033          & 17.261       & 18.904       & 24.117   & 18.072 & 22.061       & 23.073                               & \textbf{16.771}    & \textbf{17.055}                             \\ \cline{3-12} 
                                    &                                                                                   & 10 \%                                                                                   & 19.678          & 18.862          & 19.053         & 24.892      & 18.147    & 22.755       & 24.161    & \textbf{16.892}               & \textbf{17.119}                                               \\ \cline{3-12} 
                                    &                                                                                   & 15 \%                                                                                   & 20.024          & 19.381           & 21.165     & 26.883       & 20.012                        & 24.819       & 24.442  & \textbf{17.329}             & \textbf{18.673}                    \\ \hline
\end{tabular}
}
\caption{Average error scores by different approaches at different levels of Rician noises on synthetic MRI images \cite{ACRIN2020}.}
\end{table}


\begin{table*}
\centering
\begin{tabular}{|c|c|c|c|}
\hline
\multirow{2}{*}{\textbf{Methods}} & \textbf{PSNR} (dB) \cite{hore2010image} & \textbf{SSIM} \cite{hore2010image}   & \textbf{RMSE} \cite{sara2019image}   \\ \cline{2-4} 
                                  & Ideal value = high & Ideal value = 1 & Ideal value = 0 \\ \hline
CNN-RL \cite{jifara2019medical}                           & 33.881                 & 0.721            & 19.149            \\ \hline
RED-CNN \cite{chen2017ieee}                               & 35.965                 & 0.778            & 18.822           \\ \hline
K-SVD \cite{Aharonksvd2006}                                & 32.798                 & 0.726            & 19.957            \\ \hline
TV \cite{said2019total}                                & 28.252                 & 0.592            & 25.103            \\ \hline
BM3D \cite{dabov2007image}                              & 35.216                 & 0.758            & 19.231            \\ \hline
NLM \cite{manjon2008mri}                               & 29.015                 & 0.683            & 22.246            \\ \hline
ADF \cite{krissian2009noise}                               & 27.321                 & 0.637            & 24.194            \\ \hline
\textbf{Proposed 3D}                         & \textbf{39.869}                 & \textbf{0.875}            & \textbf{16.930}            \\ \hline
\textbf{Proposed 2D}                         & \textbf{37.532}                 & \textbf{0.805}            & \textbf{18.031}            \\ \hline
\end{tabular}
\caption{Average error scores by different approaches on real brain MRI images \cite{ACRIN2020}.}
\end{table*}

\begin{table*}
\resizebox{\textwidth}{!}{
\begin{tabular}{|c|c|c|c|c|c|c|c|c|c|c|c|}
\hline
\multirow{2}{*}{\textbf{Measures}}                                            & \multirow{2}{*}{\textbf{\begin{tabular}[c]{@{}c@{}}Ideal \\  value\end{tabular}}} & \multirow{2}{*}{\textbf{\begin{tabular}[c]{@{}c@{}}Poisson noise\\ levels\end{tabular}}} & \multicolumn{9}{c|}{\textbf{Methods}}                                                                                                                                                                                                                                                      \\ \cline{4-12} 
                                                                              &                                                                                   &                                                                                          & {\begin{tabular}[c]{@{}c@{}}\textbf{CNN-RL}\\\cite{jifara2019medical}\end{tabular}} & 
                                    \multicolumn{1}{l|}{{\begin{tabular}[c]{@{}c@{}}\textbf{RED-CNN}\\\cite{chen2017ieee}\end{tabular}}} &
                                    \multicolumn{1}{l|}{{\begin{tabular}[c]{@{}c@{}}\textbf{K-SVD}\\\cite{Aharonksvd2006}\end{tabular}}} &
                                    \multicolumn{1}{l|}{{\begin{tabular}[c]{@{}c@{}}\textbf{TV}\\\cite{said2019total}\end{tabular}}} &
                                    
                                    \multicolumn{1}{l|}{{\begin{tabular}[c]{@{}c@{}}\textbf{BM3D}\\\cite{dabov2007image}\end{tabular}}} &
                                    {\begin{tabular}[c]{@{}c@{}}\textbf{NLM}\\\cite{manjon2008mri}\end{tabular}} & {\begin{tabular}[c]{@{}c@{}}\textbf{ADF}\\\cite{krissian2009noise}\end{tabular}} &  \multicolumn{1}{l|}{{\begin{tabular}[c]{@{}c@{}}\textbf{Proposed}\\\textbf{3D}\end{tabular}}} &
                                    \multicolumn{1}{l|}{{\begin{tabular}[c]{@{}c@{}}\textbf{Proposed}\\\textbf{2D}\end{tabular}}} \\ \hline
\multirow{3}{*}{\begin{tabular}[c]{@{}c@{}}\textbf{PSNR} (dB)\\\cite{hore2010image}\end{tabular}} & \multirow{3}{*}{High}                                                             & 5 \%                                                                                     & 34.116          & 38.932                                & 32.904                             & 29.788                           & 37.285                             & 30.601       & 28.459       & \textbf{43.129}                           & \textbf{40.024}                           \\ \cline{3-12} 
                                                                              &                                                                                   & 10 \%                                                                                    & 33.936          & 36.974                                & 31.202                             & 29.143                           & 36.112                             & 30.212       & 28.336       & \textbf{42.263}                           & \textbf{38.774}                           \\ \cline{3-12} 
                                                                              &                                                                                   & 15 \%                                                                                    & 33.178          & 36.049                                & 31.922                             & 28.434                           & 35.117                             & 29.221       & 28.363       & \textbf{40.663}                           & \textbf{37.747}                           \\ \hline
\multirow{3}{*}{\textbf{SSIM} \cite{hore2010image}}      & \multirow{3}{*}{1}                                                                & 5 \%                                                                                     & 0.726           & 0.791                                 & 0.712                              & 0.613                            & 0.745                              & 0.692        & 0.623        & \textbf{0.802}                            & \textbf{0.798}                            \\ \cline{3-12} 
                                                                              &                                                                                   & 10 \%                                                                                    & 0.722           & 0.785                                 & 0.710                              & 0.609                            & 0.740                              & 0.689        & 0.622        & \textbf{0.800}                            & \textbf{0.792}                            \\ \cline{3-12} 
                                                                              &                                                                                   & 15 \%                                                                                    & 0.714           & 0.735                                 & 0.702                              & 0.539                            & 0.725                              & 0.661        & 0.613        & \textbf{0.797}                            & \textbf{0.751}                            \\ \hline
\multirow{3}{*}{\textbf{RMSE} \cite{sara2019image}}      & \multirow{3}{*}{0}                                                                & 5 \%                                                                                     & 19.164          & 18.762                                & 20.882                             & 26.115                           & 19.102                             & 24.421       & 26.148       & \textbf{17.948}                           & \textbf{18.030}                           \\ \cline{3-12} 
                                                                              &                                                                                   & 10 \%                                                                                    & 20.273          & 19.883                                & 21.015                             & 26.862                           & 19.911                             & 25,124       & 26.924       & \textbf{18.041}                           & \textbf{19.002}                           \\ \cline{3-12} 
                                                                              &                                                                                   & 15 \%                                                                                    & 20.913          & 20.182                                & 21.995                             & 28.834                           & 20.232                             & 26.224       & 28.363       & \textbf{18.685}                           & \textbf{19.703}                           \\ \hline
\end{tabular}
}
\caption{Average error scores by different approaches at different levels of Poisson noise on synthetic CT images \cite{ACRIN2020}.}
\end{table*}

\begin{table*}
\centering
\begin{tabular}{|c|c|c|c|}
\hline
\multirow{2}{*}{\textbf{Methods}} & \textbf{PSNR} (dB) \cite{hore2010image} & \textbf{SSIM} \cite{hore2010image}   & \textbf{RMSE} \cite{sara2019image}   \\ \cline{2-4} 
                                  & Ideal value = high & Ideal value = 1 & Ideal value = 0 \\ \hline
CNN-RL \cite{jifara2019medical}                           & 35.204                 & 0.756            & 19.013            \\ \hline
RED-CNN \cite{chen2017ieee}                               & 39.991                 & 0.801            & 18.011           \\ \hline
K-SVD \cite{Aharonksvd2006}                                & 32.778                 & 0.704            & 20.206            \\ \hline
TV \cite{said2019total}                                & 29.484                 & 0.625            & 25.020            \\ \hline
BM3D \cite{dabov2007image}                              & 38.110                 & 0.772            & 18.996            \\ \hline
NLM \cite{manjon2008mri}                               & 30.301                 & 0.699            & 24.003            \\ \hline
ADF \cite{krissian2009noise}                               & 28.212                 & 0.626            & 25.116            \\ \hline
\textbf{Proposed 3D}                          & \textbf{42.741}                 & \textbf{0.893}            & \textbf{16.557}            \\ \hline
\textbf{Proposed 2D}                          & \textbf{40.974}                 & \textbf{0.882}            & \textbf{17.831}            \\ \hline
\end{tabular}
\caption{Average error scores by different approaches on real brain CT images \cite{ACRIN2020}.}
\end{table*}
\par \textit{Setup and Parameters}: In our experiment, \textbf{(a) for $3D$ denoising}: we have used 400 slices of $3D$ MRI and $350$ slices of CT images having $256 \times 256$ voxels, each voxel is of resolution $1 mm \times 1 mm \times 1 mm$. We have added different levels of noise in MRI and CT images. Then we test our model using $50$ and $45$ slices of $3D$ MRI and CT real datasets, respectively. As shown in Fig. 1 we select one block of $3D$ data at a time and decompose it into overlapping block patches. We have a total of $64$ block patches each of dimension $32 \times 32 \times 8$ voxels within a block of $256 \times 256 \times 192$ voxels. Then each block patch is provided to the DL and RL parts in order to generate the sparse vector and the $3D$ dictionary, respectively, to reconstruct the denoised patches. The deep residue network in RL consists of $14$ layers where the first layer is a combination of $3D$ convolution and $3D$ ReLU. $3D$ convolution consists of $84$ filters of dimension $3 \times 3 \times 8$ followed by $max3D$ operation to introduce the non-linearity. Then $3D$ batch normalization is added in the next $12$ layers to uplift the denoising performance by using higher learning rates. The final layer is the $3D$ convolution layer that gives the learned residue. This is used with the residue obtained from the DL part to make an average residue $\mathbf{R}_{avg}$. The averaged residue $\mathbf{R}_{avg}$ is again fed back to update the dictionary in DL part. Note that now the learned sparse dictionary can efficiently reduce Rician noise and Poisson noise from the MRI and CT images, respectively. \textbf{(b) for $2D$ denoising}: we have used $1500$ slices of MRI and $1000$ slices of CT images. Each image has a dimension of $512 \times 512$ pixels. Then we test our model using $442$ and $250$ image slices for real $2D$ datasets of MRI and CT images, respectively. We have a total of $84$ patches each of dimension $64 \times 64$ pixels within an image of $512 \times 512$ pixels. Note that now the learned sparse dictionary can efficiently reduce Rician noise and Poisson noise from the MRI and CT images, respectively. 

\par The regularization parameters $\lambda$ and $\mu$ are chosen after performing many trials on different noise levels on $3D$ and $2D$ images for MRI and CT images. Fig. 3 shows the values of peak signal-to-noise ratio (PSNR) obtained at different range of $\lambda$ and $\mu$. As shown in Fig. 3 they are fixed to $\lambda = 0.5$ and $\mu = 1$. For implementing other comparative approaches, we use the optimal values of parameters as available in respective papers \cite{chen2017ieee, jifara2019medical, Aharonksvd2006, said2019total, dabov2007image, manjon2008mri, krissian2009noise}.

\par \textit{Result analysis}: We first display/discuss the visual results and then present the quantitative analysis with different performance metrics. Fig. 4 shows the denoising results of synthetic MRI data. We add $5 \%$ of Rician noise in the ground truth image (Fig. 4(a)) and apply different algorithms to denoise it. Fig. 4(b) shows that CNN-RL \cite{jifara2019medical} maintains the outer details but inner details are pixelated. In Fig. 4(c) RED-CNN algorithm \cite{chen2017ieee} used to denoise the image also maintains the outer part of the edges whereas the inner details are still pixelated however it has improved results than CNN-RL image. In Fig. 4(d) we see that K-SVD \cite{Aharonksvd2006} successfully preserves the edges to some extent but inner details are not clear. Fig. 4(e) shows a TV approach \cite{said2019total} is also not able to preserve the edges of the images. In Fig. 4(f) from the zoom portion, one can observe that BM3D \cite{dabov2007image} is able to maintain the outer edges but inner details of the image are appearing blur. From Fig. 4(g) one can observe that NLM \cite{manjon2008mri} is not able to maintain the sharpness in edges as they get distorted. Fig. 4(h) shows that ADF \cite{krissian2009noise} blurs the image and edge width is also increased. In Fig. 4(i) one can see that the image generated by the proposed unsupervised approach for $3D$ is nearly close to the ground truth image as well as it preserves both the inner and outer details of the image. Even the proposed approach result, Fig. 4(j) obtained for $2D$ image is also preserving the edges. The quantitative evaluation is presented in Table I where one can observe calculated values of PSNR \cite{hore2010image}, structural similarity index measure (SSIM) \cite{hore2010image}, and root mean square error (RMSE) \cite{sara2019image} of an estimated denoised image by implementing different algorithms at various noise levels in the imaging. It can be seen that the PSNR and SSIM of the denoised image estimated by both the $3D$ and $2D$ proposed approach is higher than other methods while the RMSE value is at a low when compared to other methods.

\par Fig. 5 shows the qualitative results on the real MRI dataset \cite{ACRIN2020}. In Fig. 5(a) one can observe that the result of CNN-RL \cite{jifara2019medical} is pixelated. Fig. 5(b) is the result of RED-CNN method \cite{chen2017ieee} and here too the edges are not clear and pixelated. Fig. 5(c) shows the K-SVD output \cite{Aharonksvd2006} and again the finer details are missing. In Fig. 5(d) TV \cite{said2019total} is used to denoise the image and one may see that the edges are not as clear that degrades the visual quality of image. Fig. 5(e) is the result obtained by BM3D \cite{dabov2007image} method and one can notice that outer edges are sharp, however, inner details are blur. Fig. 5(f) shows the result obtained by NLM method \cite{manjon2008mri} and one can see that edges are not clear and noise is still present in the image that decreases the visual quality of image. In Fig. 5(g) one can observe that the entire image generated by ADF method \cite{krissian2009noise} is blur and visual quality is also poor. Fig. 5(h) and 5(i) show the results obtained from the proposed unsupervised $3D$ and $2D$ learning approach and it is visible that our model is able to reduce noise to a greater extent as compared to other approaches. Table II lists average error scores obtained by different approaches on real brain MRI images. From the table, one can observe that the proposed approach has PSNR and SSIM value higher than other approaches. RMSE value of the proposed approach is low when compared to other approaches.

\par Fig. 6 shows the visual results of the denoised image reconstructed from different denoising methods applied on synthetic CT images obtained by adding $5 \%$ Poisson noise to the Shepp-logan dataset. Shepp-logan dataset is the synthetic dataset that is widely used for research purposes. In Fig. 6(b) CNN-RL \cite{jifara2019medical} maintains the structure of the image, however, due to the formation of grainy structures in the image edges are not distinct. Fig. 6(c) shows that RED-CNN \cite{chen2017ieee} reduces the noise to some level, however, the structure is shifted a bit from its original position. Fig. 6(d) shows that K-SVD \cite{Aharonksvd2006} is not able to smooth the homogeneous regions that are making the image pixelated. In Fig. 6(e) we observe that TV output \cite{said2019total} is blur and one cannot differentiate the boundaries. Fig. 6(f) indicates that overall the noise is reduced using BM3D method \cite{dabov2007image} however finer details can be improved. Fig. 6(g) is the NLM \cite{manjon2008mri} denoised output in which noise is visible. In Fig. 6(h) we see that ADF \cite{krissian2009noise} blurs the image that degrades the visual quality of the image. Fig. 6(i) shows the result obtained from the proposed unsupervised learning approach in $3D$ way and one can observe that the reconstructed image is close to the ground truth image (Fig. 6(a)) and also the Poisson noise is reduced to a great extent. The result obtained in Fig. 6(j) from proposed approach for denoising image in $2D$ method also decreases the noise level to some extent. Table III shows the respective quantitative results of all the methods for different noise levels and it is visible that PSNR \cite{hore2010image} and SSIM \cite{hore2010image} of the proposed approach are higher than other approaches and RMSE \cite{sara2019image} value of the proposed approach is at a low. 

\par Fig. 7 shows the results obtained by applying the denoising methods on real CT datasets \cite{ACRIN2020}. In Fig. 7(a) we observe that CNN-RL \cite{jifara2019medical} is not able to preserve the edges due to which one can not differentiate between the boundaries. Fig. 7(b) shows that RED-CNN \cite{chen2017ieee} maintains the edges however noise content is still present in the image. The output in Fig. 7(c) shows that the structure is maintained by K-SVD \cite{Aharonksvd2006} however details in the image are lost. Fig. 7(d) shows that the image generated by applying TV method \cite{said2019total} on the input image is not able to preserve the structure in the image. In Fig. 7(e) it is clear that BM3D \cite{dabov2007image} is able to preserve the edges however noise is still present in homogeneous regions of the image. In Fig. 7(f) one can see that the image obtained by NLM \cite{manjon2008mri} still is very noisy. Fig. 7(g) shows that ADF \cite{krissian2009noise} makes the image blur thus any part of the image is not clearly visible. In Fig. 7(h) and Fig. 7(i) one can observe that proposed unsupervised learning approach performs better than other existing seven state-of-the-art approaches by preserving the edges and maintaining the visual quality. Table IV shows the quantitative results between the existing state-of-the-art approaches and the proposed approach for real CT datasets. From both Fig. 7 and Table IV, one can see that the proposed approach following the unsupervised deep learning concept performs better than the other approaches.

\section{Conclusion and Future work}

We have presented a novel unsupervised deep learning approach for medical image denoising considering input as $2D$ and $3D$ for image/voxel processing. The proposed framework takes care of both the Rician noise and Poisson noise present in the MRI and CT images, respectively. Our model learns the patch-based dictionaries in order to learn noise indirectly while it learns the residue (noise) contents directly from the available MRI/CT images using proposed deep residue network. Note that the proposed approach does not require the clean (denoised) images for training the model, unlike many deep learning-based recent approaches. We have better handled the ill-posed nature of the problem by choosing the optimum regularization parameters that we have estimated from the data. Dictionary-based deep residue network reduces the noise from the images by preserving the edges of the images and maintaining their visual quality (without losing details) which is evident from the results. In future, we would like to work on the restoration of images and overcome any degradation in the medical imagery along with the noise. 

\section*{Acknowledgment}
The authors would like to thank the editors, the associate editor, and anonymous reviewers for their insightful and helpful comments. We are thankful to the Gujarat Council on Science and Technology (GUJCOST) for providing PARAM Shavak GPU-based supercomputer to our institute and allow us to conduct the exhaustive experiments in this research work. 


\bibliographystyle{IEEEtran}
\bibliography{ref}


\end{document}